\documentclass[english,aps,prc, twocolumn,nofootinbib,0superscriptaddress]{revtex4-1}
\usepackage{epsfig}
\usepackage{graphicx}
\usepackage{amsmath}
\usepackage{lipsum}
\usepackage{color}


\def\orcid#1{\kern .08em\href{https://orcid.org/#1}{\includegraphics[keepaspectratio,width=0.7em]{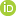}}}

\usepackage{babel}

\usepackage[colorlinks,linkcolor=blue,anchorcolor=blue,citecolor=blue,urlcolor=blue,filecolor=black]{hyperref}

\newcommand{\beq}{\begin{equation}}
\newcommand{\eeq}{\end{equation}}
\newcommand{\bea}{\begin{eqnarray}}
\newcommand{\eea}{\end{eqnarray}}
\usepackage{graphicx}

\begin{document}


\title{Probing the size and binding energy of the hypertriton in heavy ion collisions}

\author{C.A. Bertulani \!\! \orcid{0000-0002-4065-6237}}
\email[]{Email: carlos.bertulani@tamuc.edu}
\affiliation{Department of Physics and Astronomy, Texas A\&M University-Commerce,  TX 75429-3011, USA}

\begin{abstract}
The hypertriton is predicted to have a small binding energy (a weighted average of about 150 keV), consistent with
a large matter radius ($\sim 10$ fm), larger than the historical $^{11}$Li halo discovered more than 35 years ago. But
the reported experimental values of the binding energy of the hypertriton range from 70 to 400 keV. In this work I
discuss the electromagnetic response and interaction radius of the hypertriton and how high energy heavy
ion collisions ($\sim 1 - 2$ GeV/nucleon) can help achieving a higher accuracy for the determination of its size and binding
energy.

\end{abstract}

\maketitle

{\bf Introduction.} The equation of state (EOS) governing the structure of a neutron star (NS) is one of the main focus of recent nuclear physics studies, invoking new theoretical methods, dedicated experiments at nuclear physics facilities, and astronomical observations with advanced telescopes and satellites \cite{Lattimer-2021}. The EOS is mainly ruled by the strong interaction within the neutron matter, but there is a consensus in the nuclear theory community that hyperons may appear in the inner core at about twice the nuclear saturation density (the nuclear saturation density is $\rho_0 = 0.16$ fm$^{-3}$) \cite{RevModPhys.88.035004,TOLOS2020103770}. The presence of hyperons in the medium reduces the nucleonic Fermi energy and consequently the nuclear EOS is softened, limiting the mass that a NS can reach. This expectation has been challenged by the unexpected observation of NSs larger than twice the solar mass \cite{Demorest-2010,Barr-2016,Cromartie2020,Fonseca-2021}. This conundrum has been termed as the ``hyperon puzzle", although there are still some EOS that include hyperons but can still  stabilize heavy NSs.

Key to understand the role of hyperons in NS is the hyperon-nucleon interaction. A plethora of nuclear experiments have been performed or are being planned to produce and study the properties of ``hypernuclei'', a term typically used for a $\Lambda$ particle imbedded in a nucleus. Numerous properties of $\Lambda$-hypernuclei have been reported and compared to theoretical expectations. But none of these properties is more important than their binding energy.  Its determination allows the theoretical adjustment of the $\Lambda$-nucleon interaction and to further predict other properties such as excitation energies, sizes and lifetimes.  All these quantities, including the spins of light hypernuclei, are expected to be closely related \cite{DALITZ196258,Rayet-1966}, as observed in the case of halo nuclei \cite{TanihataPRL.55.2676}.

\begin{figure}[t]
\begin{center}
\includegraphics[
width=3.5in
]
{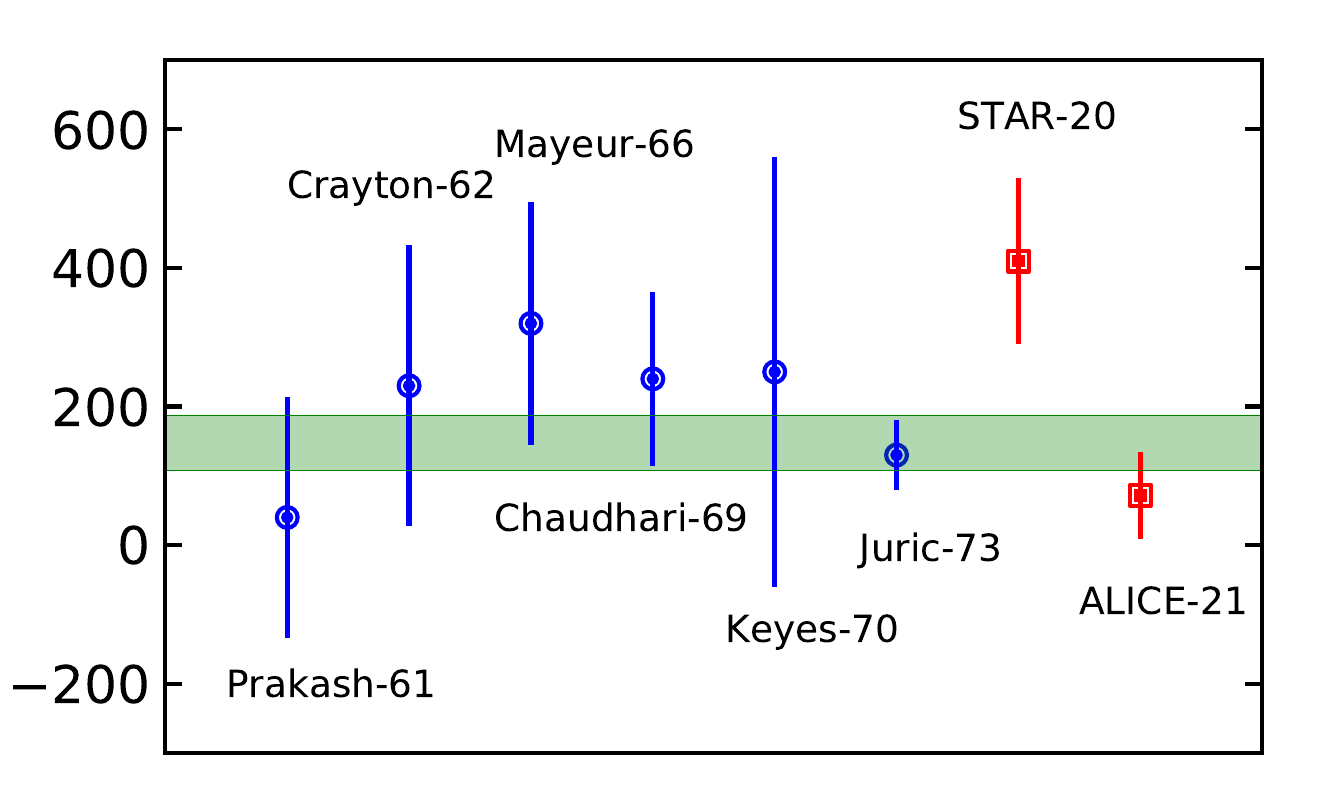}
\caption{Measurements of the binding energy of the hypertriton obtained with two-body and three-body pionic decays in emulsion and in heavy ion reaction experiments.  The mean value was evaluated in Ref. \cite{MainzDataBase} to be $B_\Lambda = 148 \pm 40$ keV (shaded band with statistical and total uncertainties). }
\label{BL}
\end{center}
\end{figure}

The lightest observed hypernucleus is the hypertriton  $^3_\Lambda$H, which has a large binding energy experimental uncertainty. Being the simplest of all hypernuclei, it is imperative to decrease this experimental uncertainty allowing for a more reliable description of heavier hypenuclei. A compilation of the $^3_\Lambda$H binding energy  measurements is shown in Figure \ref{BL}, with data extracted from Refs. \cite{Prakash1961,RevModPhys.34.186,Mayeur1966,Chaudhari1968,BOHM1968511,JURIC19731,DAVIS20053,AdamNature2020}. For simplicity, I assume that  statistical and systematical errors  are uncorrelated and added according to $\sigma^2 = \sigma_{stat}^2 + \sigma_{sys}^2$. While  Ref. \cite{JURIC19731} reports $B_\Lambda = 130 \pm 50$ keV,  a recent STAR measurement obtained $410 \pm 120$  keV \cite{AdamNature2020}. An ALICE experiment   reported a much smaller  value of $72^{+63}_{-36}$ keV   \cite{arXiv:2209.07360}.
The shaded band shown in Figure \ref{BL} is the result of a statistical model analysis conducted in  Ref. \cite{MainzDataBase} for the average and uncertainty of the binding energies based on experimental errors and significant weights. The analysis  of all existing data suggests that the $\Lambda$ is bound to a deuteron by only $148 \pm 40$ keV. 

The small binding energy in $_3^\Lambda$H  implies that it has an extended wave function for the $\Lambda$ with respect to the deuteron core, which has a much larger binding energy $B_d \simeq 2.2$ MeV. Thus the $\Lambda$  forms a ``halo" system with an approximate size $R = \sqrt{\hbar^2/(4\mu B_\Lambda)} \simeq 10 $ fm, where $\mu$ is the reduced mass and $B_\Lambda$ the hypertriton binding energy \cite{ISI:A1993LF28300001}. This is probably the largest known halo nucleus since the discovery of the $^{11}$Li halo nucleus about 35 years ago \cite{TanihataPRL.55.2676}. 

The binding energy of the hypertriton is often used as a benchmark for the $\Lambda$N interaction, which is an important constraint in attempts to resolve the ``hyperon puzzle'' related to whether or not dense NSs  contain hyperons, besides nucleons \cite{Lattimer-2021,RevModPhys.88.035004,TOLOS2020103770}. One has also been invoked a three-body $\Lambda$NN interaction together with a two-body interaction to study the hypertriton binding and other properties of hypernuclei \cite{Gal_2020,kohno2022partial}. The conclusion of these works is that the $\Lambda$NN interaction forces are not testable by studying the hypertriton. $B_{\Lambda}$ is used in most YN studies to constrain the (S=1)/(S=0) ratio of the $\Lambda$-N 2-body force. Other studies have explored the details of the $\Lambda$N  interaction  using three-body models and effective field theoretical (EFT) techniques \cite{Cobis-1997,PRC.100.034002,friedman2022constraints}. To further elucidate these questions, experimental studies of the hypertriton size and its corresponding binding energy  have been proposed by measuring  the interaction cross section of the hypertriton produced in  reactions with heavy ion beams \cite{PBM2019,obertelli22}.

\begin{figure}[t]
\begin{center}
\includegraphics[
width=3.in
]
{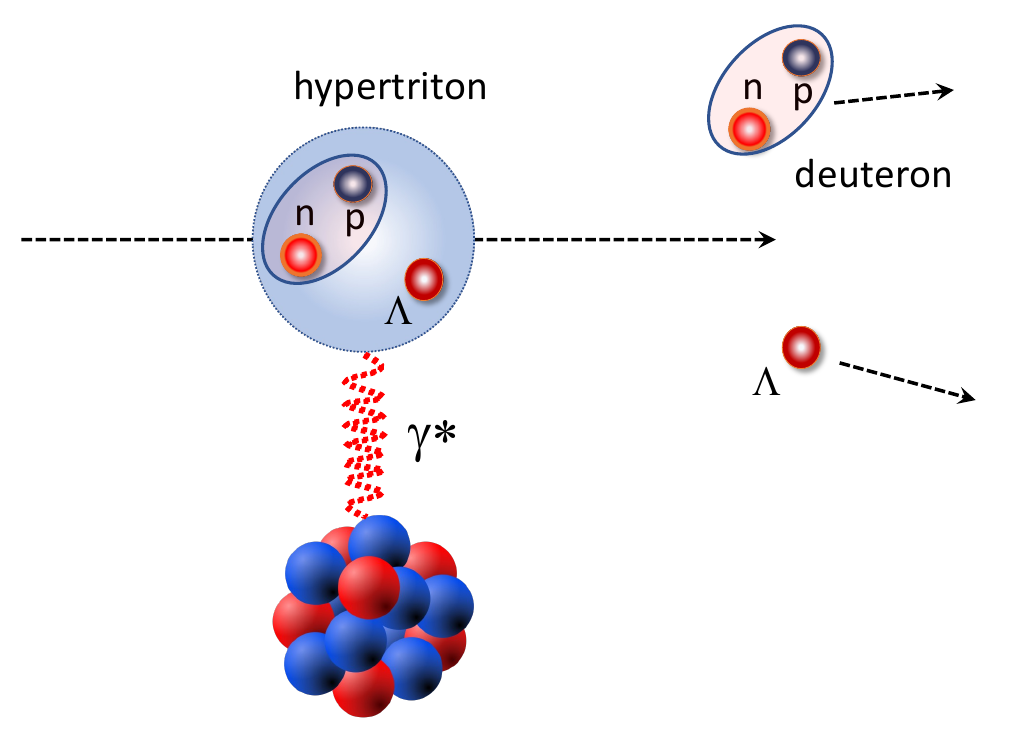}
\caption{Schematic view of the dissociation of the hypetriton  in the Coulomb field of a nuclear target.}
\label{CBL}
\end{center}
\end{figure}

In this work, I explore how the hypertriton size and binding energy affects the magnitude of the electromagnetic breakup and interaction cross sections for hypertritons imping on carbon, tin and  lead active targets (see Figure \ref{CBL}). As a benchmark, I consider the reaction $^3_\Lambda$H + A $\rightarrow$ $\Lambda$ + d + A at 1.5 GeV/nucleon with A = C, Sn and Pb targets. In particular, I show that the electromagnetic response of the hypertriton is very sensitive to its binding energy, yielding large breakup cross sections, comparable with the nuclear interaction cross sections. The electromagnetic response and the interaction cross section adds value to each other and can be complementarily used as standard probes of the binding energies and sizes of loosely-bound hypernuclei.  

{\bf Theoretical framework.} I will consider the general dependence of the electromagnetic response and interaction cross sections of the hypertriton with different target nuclei. At this time, there are no experiments to compare with and the exact details of the hypertriton wavefunction are only qualitatively relevant. I therefore adopt a simplified model for the structure of the hypertriton, treating it as a $\Lambda$ particle bound to a deuteron  by means of a Woods-Saxon  (WS) potential. Similar models were proven very useful to understand the location of binding energies and single particle states in numerous hypernuclei along the periodic table \cite{PRC.38.2700,RevModPhys.88.035004}. 

For the electromagnetic breakup of the hypertriton in the active target I assume an electric dipole transition from the s-wave ground state ($l_0=0$) to the continuum. The transition $\gamma + _\Lambda^3$H $ \rightarrow d + \Lambda$, from the ground state (g.s.)  of $_\Lambda^3$H  to a continuum of $d + \Lambda$ with partial wave quantum numbers $l$, is considered. The bound-state wavefunctions are normalized to unity, $\int dr \left\vert
u_{0}\left(  r\right)  \right\vert ^{2}=1$, whereas the continuum
wavefunctions for the $\Lambda$-d system have boundary conditions at infinity given by $-\sqrt{2\mu_{\Lambda d}/\pi\hbar^2k} e^{i\delta_l}\sin(kr + \delta_l)$, where  $k$ is the wavenumber and $\delta_l$ is the partial wave phase-shift. With these definitions, the continuum
wavefunctions are properly normalized as $\left\langle
u_{E^\prime}|u_{E} \right\rangle =\delta\left(
E^\prime-E\right)  \delta_{l l^\prime}. $
The E1 operator will constrain transitions from the ground state to $p$-wave continuum states.

The multipole strength, or response function,  for this system is given by \cite{Bertulani1996}
\begin{equation}
{dB(E1; g.s.\rightarrow El)\over dE} = {1\over \hbar} \sqrt{\mu \over 2E}{\left| \left< g.s. \|{\cal O}_{E1} \|  El \right> \right|^2 }, \label{dbdk}
\end{equation}
where $E$ is the relative kinetic energy of the $\Lambda$ + deuteron, and $\left\langle
g.s.\left\Vert \mathcal{O}_{E1}\right\Vert El \right\rangle$ is the dipole matrix element 
\begin{align}
\left\langle g.s.\left\Vert
\mathcal{O}_{E1}\right\Vert El \right\rangle
&=(-1)^{l}{e_{eff} \over
\sqrt{4 \pi}}
\int_{0}^{\infty}dr \,
ru_{g.s.}(r)u_E(r)
,\label{lol0}%
\end{align}
with  $u_{g.s.}$ ($u_E$) being the ground-state (continuum) wavefunction, obtained from the solution of the radial Schr\"odinger equation, and the effective charge $e_{eff}=e(m_\Lambda/m_{^3_\Lambda{\rm H}})$ which accounts for the center-of-mass motion. 

The virtual photons provided by the interaction of the target with the hypertriton will lead to its Coulomb dissociation, with the differential  cross section
\begin{equation}
{d\sigma_C \over  dE} = {16\pi^3  \over 9\hbar c }  n(E){dB(E) \over dE} , \label{dsCoulde}
\end{equation}
where  the equivalent photon  numbers are given by \cite{BERTULANI1993158}
\begin{equation}
n(E) = {2Z_T^2\alpha \over \pi} \left( {Ec\over \gamma \hbar v^2}\right)^2  \int_0^\infty db \, b \left[ K_1^2 +{1 \over \gamma^2}K_0^2\right] T(b), \label{equiv}
\end{equation}
with $Z_T$ being the projectile charge, $\alpha$ the fine-structure constant, $v$ the projectile (hypertriton) velocity, $\gamma = (1-v^2/c^2)^{-1/2}$ is the Lorentz contraction and the modified Bessel functions $K_n$ have the argument $x=Eb/\gamma\hbar v$. 

The transparency function appearing in Eq. \eqref{equiv}  accounts for absorption at small impact parameters, i.e., the probability that the hypertriton survives a collision with the target. It can be calculated using eikonal waves and the Glauber multiple scattering theory based on binary hadronic collisions. The transmission is a product of the survival probabilities of $\Lambda$,  proton and neutron interacting with the target, i.e.,  
\beq
T(b) = T_\Lambda (b) T_p(b) T_n(b),
\eeq 
where
\bea
T_i(b)    &=&
\int d^{2}s_i\;dz_i \;\rho_i\left(  z_i,\mathbf{s_{\rm i} -b}\right)
 \nonumber \\ 
&\times&\exp\left[  -\sigma_{pi}Z_T\int dz'\;\rho_p^{T} \left(z',\mathbf{s}\right)\right] \nonumber \\
&\times&  \exp\left[-\sigma_{ni}N_T \int dz'\;\rho_n^{T}\left(z',\mathbf{s}\right)  \right] , 
 \label{abrasionp2}
\eea
with $i=\Lambda$, $p$, or $n$, and all densities normalized to unity. 

The proton and neutron densities in $_\Lambda^3$H are assumed to be the same and are obtained from a calculation  for the  deuteron wavefunction using the Av18 interaction \cite{PhysRevC.51.38}. The deuteron density is the outcome of an incoherent sum of the square of the S and D wavefunctions in its ground state. The proton and neutron densities within $_\Lambda^3$H  are calculated from the free deuteron density, $\rho_d({\bf r})$, obtained in this fashion. The proton and neutron densities are calculated with respect to the  center of mass of the deuteron, where I assume equal neutron and proton masses, $m_p=m_n=m_N=939$ MeV. For more details, see supplement.

The $\rho_\Lambda({\bf r}_\Lambda)$ single-particle density is obtained from the square of the wavefunction resulting from the solution of the $\Lambda + d$  Schr\"odinger equation with the WS potential model described above. The neutron and $\Lambda$ densities need to account for the center of mass displacement of the calculated wavefunctions. In the supplemental material I describe how the densities and S-matrices are calculated. After the center of mass corrections are performed, the nucleon and $\Lambda$ densities are normalized to unity to comply with Eq. \eqref{abrasionp2}.
  
The total reaction cross section is given by the sum $\sigma_R = \sigma_C+ \sigma_I$ of the Coulomb breakup cross section, $\sigma_C = \int dE \, d\sigma_C /  dE$ and the ``interaction cross section"
\begin{equation}
{\sigma_I } = 2\pi \int db \, b \left[1-T(b)\right]. \label{sigI}
\end{equation}

\begin{figure}[t]
\begin{center}
\includegraphics[
width=3.5in
]
{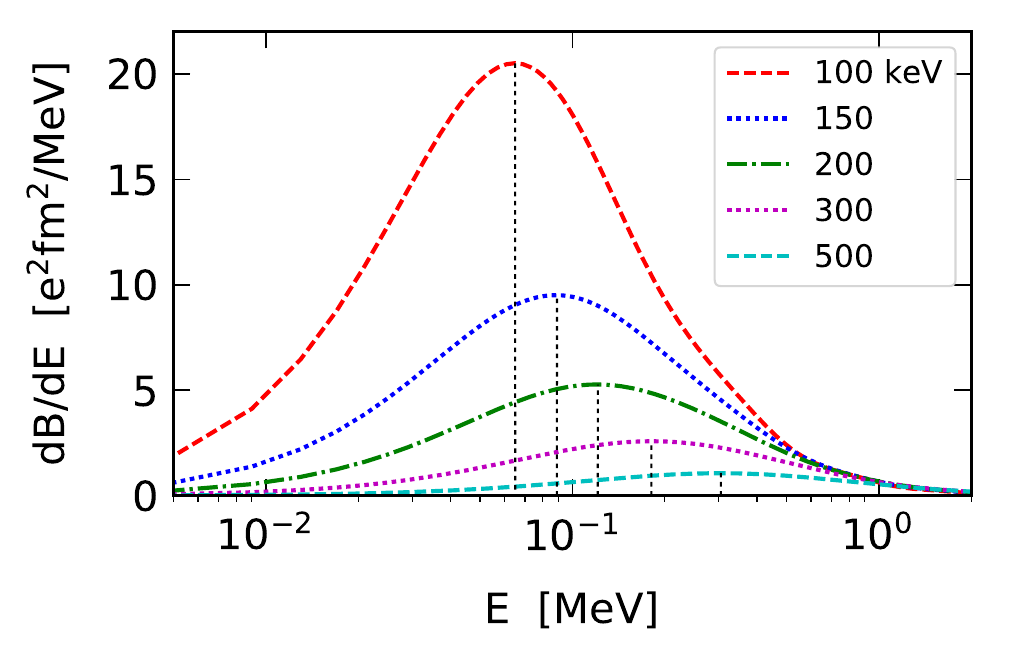}
\caption{Electric dipole response function, $dB/dE$, for $_\Lambda^3$H as a function of the relative kinetic energy of the final state of $\Lambda$-d. The calculations are performed for  $\Lambda$-d separation energies  $B_\Lambda$ equal to 100, 150, 200, 300, and 500 keV. The vertical lines are analytical predictions for the maxima of the response functions from Ref. \cite{bersus1992}.}
\label{response}
\end{center}
\end{figure}

{\bf Results.} 
WS models have been used with success previously to reproduce binding energies and single particle states of  hypernuclei yielding insights on their additional properties \cite{PRC.38.2700,RevModPhys.88.035004}. In this work,
I calculate the bound state (s-wave) and continuum (p-wave) wavefunctions of the $_\Lambda^3$H system using a WS potential with parameters $R= 2.5$ fm (range) and $a=0.65$ fm (difuseness). Its depth was adjusted to reproduce the $\Lambda$-d separation energy values $B_\Lambda$ equal to 100, 150, 200, 300, and 500 keV. The depth and other parameters are taken to be the same for both the bound and continuum states.  The potential depths are listed in Table \ref{tab0}. The response functions $dB/dE$ in units of $e^2$fm$^2$/MeV calculated with Eq. \eqref{dbdk} are shown in figure \ref{response} as a function of the relative kinetic energy of the fragments, $E$.

\begin{table}[ht]
\begin{center}
\caption{Potential depths of the WS potential reproducing the hypertriton binding energy $B_\Lambda$. \label{tab0}}
\begin{tabular}{|c|c|c|c|c|c|c|}
\hline\hline
$B_\Lambda$ (keV) & 100 & 150 & 200 & 300 & 500        \\ 
\hline 
$V_0$ [MeV]&-11.8 & 12.2 & -12.6  & -13.2& -14.3  \\
\hline
\hline
\end{tabular}
\end{center}

\end{table}

It is instructive to compare the numerical results for the response function obtained here with the analytical model developed in Ref. \cite{ISI:A1988N117600010,bersus1992} using a Yukawa function for the bound-state wavefunction and a plane wave for the continuum. The model predicts a response function with an energy dependence   \cite{ISI:A1988N117600010,bersus1992} 
\begin{equation}
{dB \over dE} = {\cal C} \sqrt{B_\Lambda} {E^{3/2}\over (E+B_\Lambda)^4},\label{dBan}
\end{equation} 
where $B_\Lambda$ is the binding energy and ${\cal C}$ is a constant. The predicted maximum of this response function occurs at $E_{max}=3B_\Lambda/5$. The location of these maxima are shown in Fig. \ref{response} as vertical dotted lines and match perfectly the maxima of the calculated response functions using the WS model. The agreement is excellent for all binding energies shown in the figure. The shape of the response function are also very well reproduced with the analytical form of Eq. \eqref{dBan} with proper choices of the normalization constant ${\cal C}$. The differences are not larger  than 4\% for all binding energies considered here and low excitation energies, $E\lesssim 3 B_\Lambda$. 

The comparison with the analytical model of Eq. \eqref{dBan}  implies that most details of the electromagnetic response of loosely bound systems are solely governed by the external part of the bound state wavefunction. It also means that the final state distortion through the interaction between the fragments $\Lambda$ and d via the WS potential, included in the numerical calculation of the continuum waves, are rather small. The final state interaction of the fragments with the target nucleus has not been considered here, and is assumed to be small at GeV/nucleon bombarding energies.  On the other hand, the magnitude and large energy tail of the response function is sensitive to some of the details of the bound state wavefunctions, such as the asymptotic normalization coefficients, and the numerical results are not quite reproduced by the analytical formula of Ref. \cite{bersus1992}.  Recent works have proved this assertion by using ab initio wavefunctions for light nuclei in the calculation of nuclear fragmentation observables \cite{PhysRevC.104.L061602,PhysRevC.105.024613}. 

\begin{figure}[t]
\begin{center}
\includegraphics[
width=3.in
]
{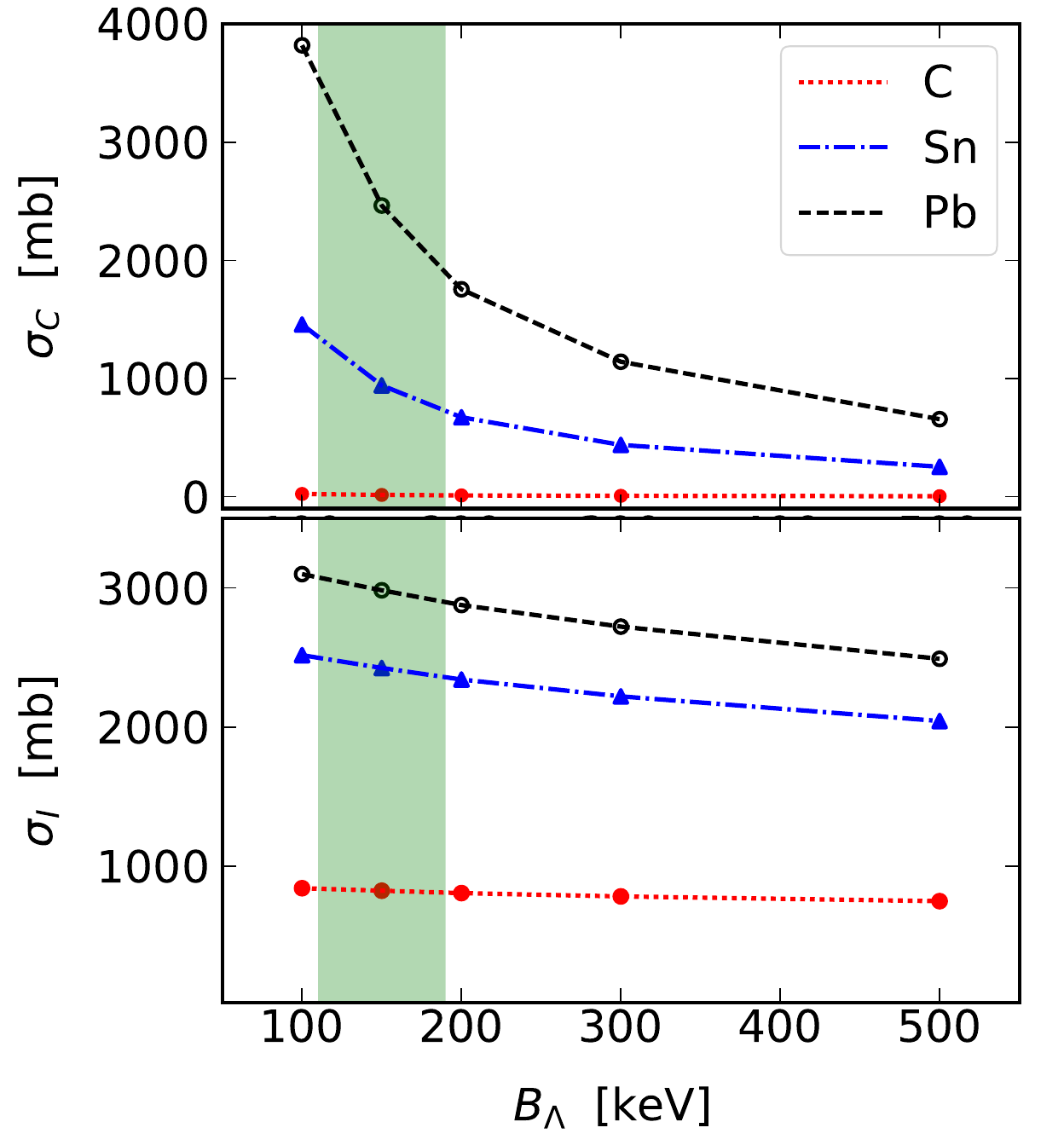}
\caption{{\it Upper panel.} Coulomb fragmentation cross section of 1.5 MeV/nucleon hypertriton projectiles impinging on carbon, tin and lead targets as a function of the hypertriton binding energy. {\it Lower panel.}  Interaction cross section of 1.5 MeV/nucleon hypertriton projectiles incident on carbon, tin and lead targets as a function of the hypertriton binding energy. The shaded bands mark the one-sigma boundaries of the average value of hypertriton binding energy ($B_\Lambda = 148 \pm 40$ keV).}
\label{coulfrag}
\end{center}
\end{figure}

\begin{table}[ht]
\begin{center}
\caption{Coulomb breakup $\sigma_C$ and interaction cross sections $\sigma_I$ for the fragmentation of 1.5 GeV/nucleon $_\Lambda^3$H impinging on $^{12}$C, $^{120}$Sn and $^{208}$Pb targets. \label{tab1}}
\begin{tabular}{|c|c|c|c|c|c|c|c|c|}
\hline\hline
$B_\Lambda$ (keV) & $\sigma_C({\rm C})$ & $\sigma_C({\rm Sn})$ & $\sigma_C({\rm Pb})$ & $\sigma_I({\rm C})$ & $\sigma_I({\rm Sn})$ & $\sigma_I({\rm Pb})$       \\ 
\hline 
100 & 22.9 & 1457.  & 3820.& 842. & 2516. &  3098.\\
\hline 
150 & 14.9 & 942.  & 2464.  & 824. &  2424. & 2982.\\
\hline
200 & 10.7 & 672.  & 1755.& 807.  & 2341. &  2876.\\
\hline
300 & 7.1 &  438. & 1142.  & 783.    &   2220. & 2721. \\
\hline
500 & 4.1 &  253.  &  656.  & 749. &   2043. &   2490.\\
\hline
\hline
\end{tabular}
\end{center}

\end{table}

In the final part of this manuscript I will discuss the numerical results for the interaction cross sections obtained for 1.5 GeV/nucleon $_\Lambda^3$H  incident on $^{12}$C, $^{120}$Sn and $^{208}$Pb targets. The calculations depend on the experimental values of the nucleon-nucleon and lambda-nucleon cross sections entering Eq. \eqref{abrasionp2}. I adopt the free pp and pn   cross sections compiled by the particle data group \cite{10.1093/ptep/ptaa104}, namely  45.8 mb  and  40 mb, respectively. For the $\Lambda$N cross section I use $\sigma_{\Lambda N} = 35$ mb, consistent with the value reported in the literature \cite{PhysRev.160.1239,GJESDAL1972152}. The numerical results for the electromagnetic breakup (upper panel) and interaction (lower panel) cross sections are shown in Fig. \ref{coulfrag}.  The shaded bands in the figure mark the one-sigma boundaries of the average value of hypertriton binding energy ($B_\Lambda = 148 \pm 40$ keV). Additionally, the numerical values for both fragmentation modes are presented in Table \ref{tab1} as a function of the hypertriton binding energy $B_\Lambda$. 

Is is visible in Figure \ref{coulfrag} as well as in Table \ref{tab1} that the Coulomb breakup cross sections for carbon targets  are much smaller than the interaction cross sections. However, they increase rapidly with the charge of the projectile and for tin and lead targets they become comparable to the interaction cross sections. It is also worthwhile noticing that the Coulomb breakup cross section has a stronger dependence with the binding energy than the interaction cross section. The advantage of interaction cross sections is that they are relatively easy to measure. For a carbon target they imply a 12\% reduction of the cross section $\sigma_I$ from $B_\Lambda=100$ MeV to $B_\Lambda=500$ MeV. This sensitivity raises to 24\% if one uses active lead targets. The sensitivity prospects increase dramatically in the Coulomb dissociation case, becoming 22\% and 600\%, respectively, for the same targets. The response function can be mapped experimentally by measuring the invariant mass of the fragments with the energies and momenta selection of the $\Lambda$ and the deuteron.   

{\bf Conclusions.}
In this work I have considered the viability  of using active targets to asses information on the binding energy of the hypertriton. These are important experiments, perhaps yielding precious information to help constrain theoretical models for the hypertriton. 

I have used a simplified model for the hypertriton wavefunctions and densities because at the pre-experiment  stage there is no need for a more elaborate calculation. The Glauber-reaction model is appropriate for the bombarding energy considered. The conclusions are that the electromagnetic response of the hypertriton is more sensitive to the binding energy than the interaction cross sections.   By looking at inclusive quantities such as the reaction vertex, and invariant mass and momenta of the fragments, a mapping of the electrogmagnetic response of the hypertriton would be of great experimental and theoretical value for the hypernuclei research community, as it has been for the study of weakly-bound nuclei.   

\medskip

\begin{acknowledgments}
{\bf Acknowledgments.} The author acknowledges support by the U.S. DOE grant DE-FG02-08ER41533 and the Helmholtz Research Academy Hesse for FAIR. 
\end{acknowledgments}


\section{Supplemental material}

\subsection{Deuteron and $\Lambda$-deuteron wavefunctions}

\begin{figure}[b]
\begin{center}
\includegraphics[
width=3.in
]
{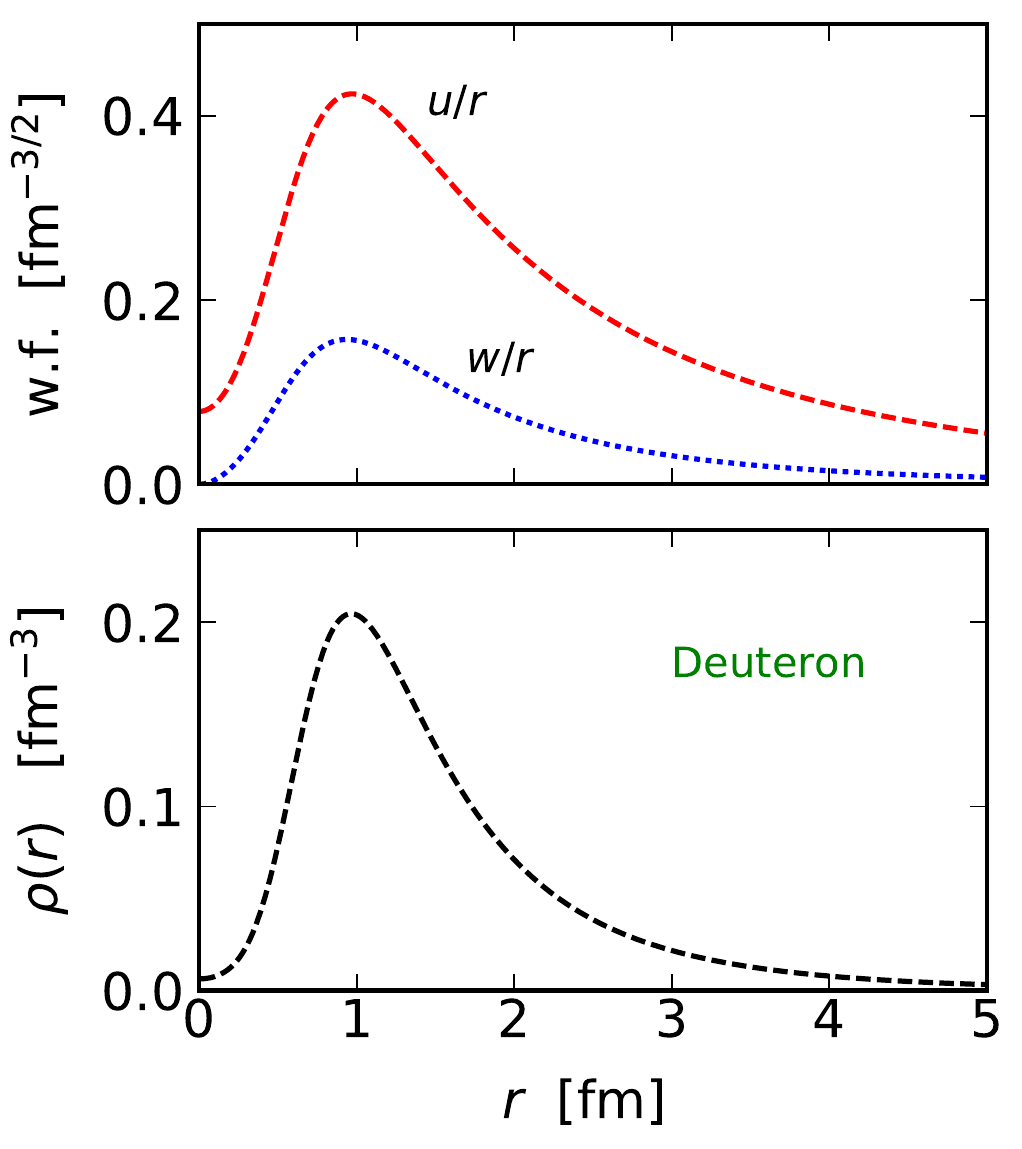}
\caption{{\it Upper panel:} Deuteron radial S-wave, $u(r)/r$, and D-wave, $w(r)/r$ wavefunctions. {\it Lower panel:} Deuteron density as a function of the proton-neutron distance $r$.}
\label{deutdens}
\end{center}
\end{figure}

The deuteron radial S-wave, $u(r)/r$, and D-wave, $w(r)/r$, wavefunctions and their densities are plotted in Figure \ref{deutdens} as a function of the proton-neutron distance $r$. The wavefunctions have been calculated with the  Av18 interaction \cite{PhysRevC.51.38} which fits the low energy pp and pn scattering data with great precision in the energy region 1-350 MeV.

\begin{figure}[t]
\begin{center}
\includegraphics[
width=3.4in
]
{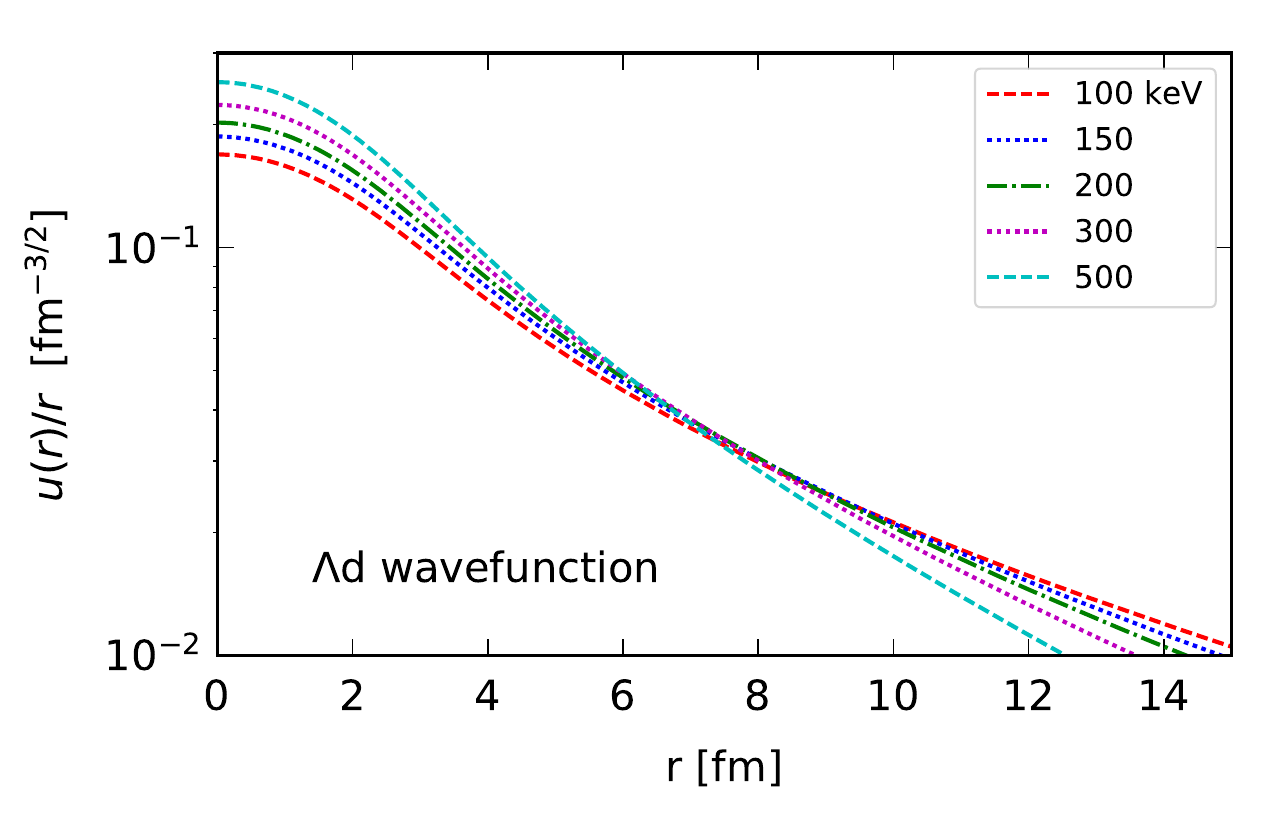}
\caption{Lambda-deuteron radial S-wave, $u_{\Lambda d}(r)/r$, as a function of the $\Lambda$-d distance $r$.}
\label{ldwf}
\end{center}
\end{figure}

The $\Lambda$-deuteron wavefunction is calculated using a Woods-Saxon (WS) potential with  a radius parameter $R=2.5$ fm and diffuseness  $a=0.65$ fm. The potential depth is chosen to reproduce the binding energy. The wavefunctions $u(r)/r$ for the S-wave are shown in Fig. \ref{ldwf} for binding energies $B_\Lambda = 100,$ 150, 200, 300 and 500 keV.

\subsection{Nucleon and $\Lambda$ densities in the hypertriton}

\begin{figure}[t]
\begin{center}
\includegraphics[
width=2.in
]
{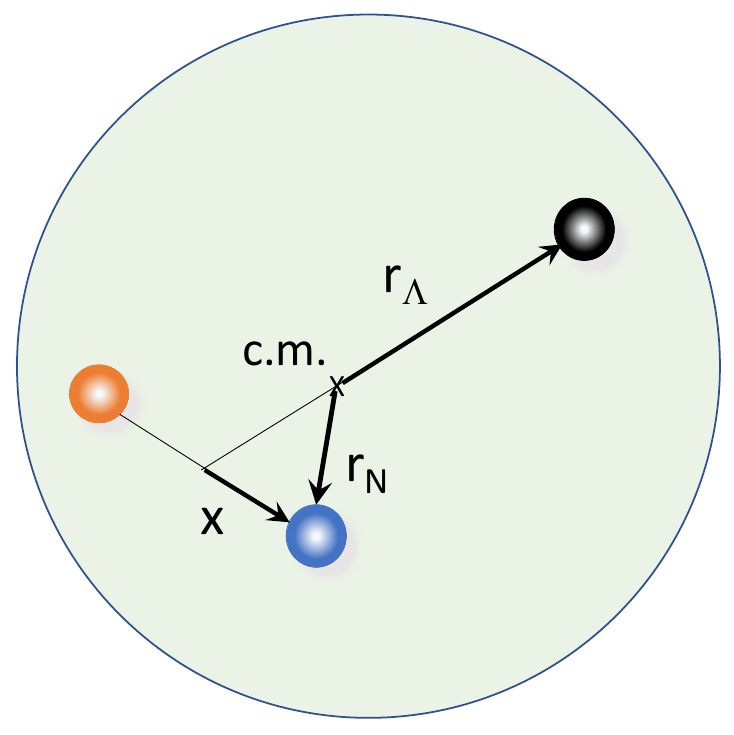}
\caption{Coordinates used in the text.}
\label{cm}
\end{center}
\end{figure}

\begin{figure}[h]
\begin{center}
\includegraphics[
width=3.5in
]
{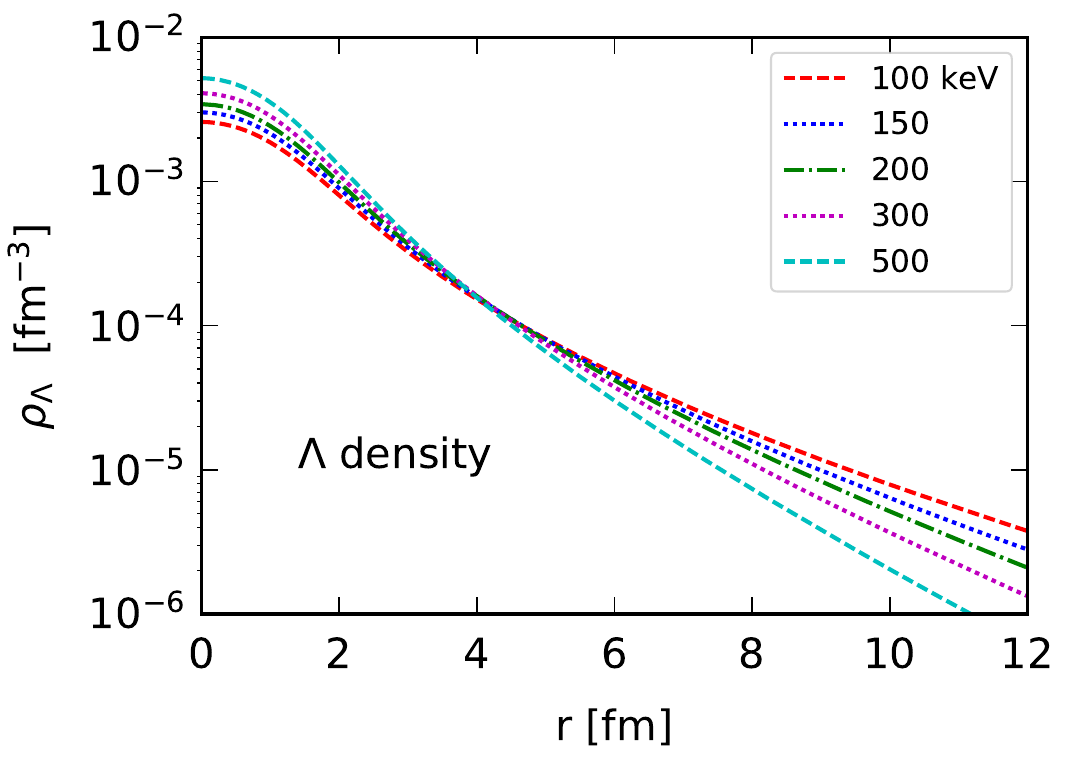}
\caption{$\Lambda$ density as a function of the distance to the c.m. of the hypertriton. }
\label{ldens}
\end{center}
\end{figure}

To calculate the nucleon and $\Lambda$ coordinate distribution within in the hypertriton with the densities and wavefunctions described above one needs to consider the center of mass correction.  The center of mass of the system also needs to be considered when determining the  nucleon densities, assumed to be the same for protons and neutrons. For clarity, I show in Figure \ref{cm} the coordinates used in the following equations. 

Defining the factors 
\beq 
\alpha = {m_d\over (m_\Lambda + m_d)} \ \ \  {\rm and} \ \ \ \beta = {m_ \Lambda\over (m_\Lambda + m_d)},
\eeq
the  $\Lambda$ density within the hypertriton as a function of its distance to the center of mass, ${\bf r}$, is given by 
\beq
\rho_\Lambda({\bf r})=|\Psi_{\Lambda d} ({\bf r}/\alpha)|^2,
\eeq 
with the previously calculated radial wavefunction  $\Psi_{\Lambda d}(r_{\Lambda d})=u_{\Lambda d}(r_{\Lambda d})/r_{\Lambda d}$. As required by the use of Eq. \eqref{abrasionp2}, the density  $\rho_\Lambda({\bf r})$ is subsequently normalized to unity.

To obtain the proton (neutron) density within the hypertriton we need to consider the probability to find the deuteron within the hypertriton multiplied by the probability to find the proton (neutron) within the deuteron. The former is given by $\rho_1({\bf r})=|\Psi_{\Lambda d} ({\bf r}/\beta)|^2$ and the later is given by $\rho_2({\bf x})=|\Psi_{deut} (2{\bf x})|^2$. 

The nucleon density at position ${\bf r}$ within the nucleus is calculated with the convolution
\beq
\rho_N({\bf r}) = \int d^3 x \rho_1(|{\bf r} - {\bf x}|)\rho_2({\bf x}),
\eeq 
This integral has cylindrical symmetry, requiring only two integrations.  As with the case of the lambda density, the nucleon density $\rho_N(r)$ is normalized to unity, in accordance with Eq. \eqref{abrasionp2}.

\begin{figure}[htb]
\begin{center}
\includegraphics[
width=3.5in
]
{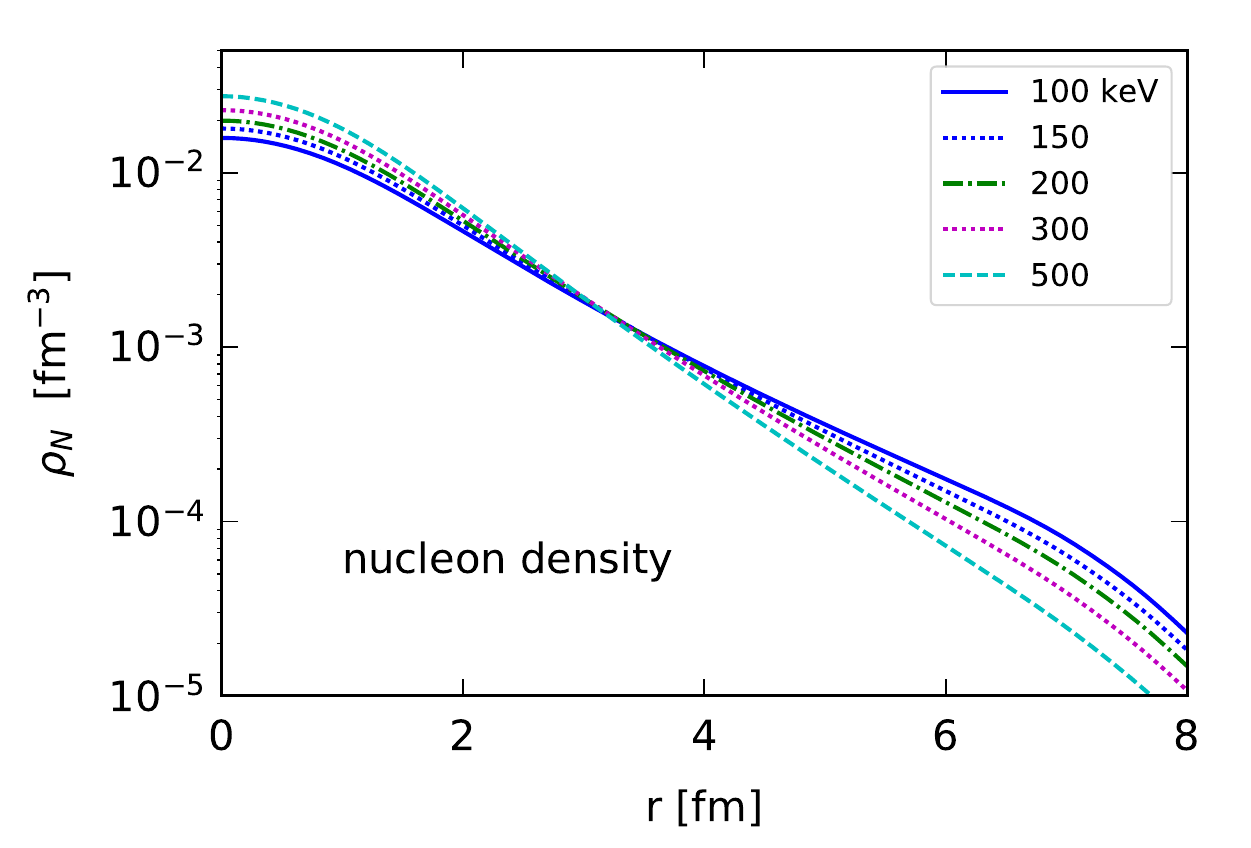}
\caption{Nucleon density as a function of the distance to the c.m. of the hypertriton. }
\label{ldens2}
\end{center}
\end{figure}

Figure \ref{ldens} shows the density distribution of the $\Lambda$ within the hypertriton as a function of its distance to the c.m. of the system.  As expected for a S-wave, the density is peaked at small distances and decays linearly at larges distances, with a larger slope for larger binding energy.   The nucleon (proton or neutron) density in the hypertriton is displayed in Figure \ref{ldens2}  as a function of its distance to the c.m. of the system.  A similar behavior is observed, as in the case of the $\Lambda$ distribution. But in contrast to it, there is a shaper decrease of the nucleon distribution for large distances due to the stronger binding of the nucleons within the deuteron inside the hypertriton. At the center of the hypertriton, it is more likely to find a nucleon than the $\Lambda$ particle by a factor of 10, as the $\Lambda$ spreads out to much larger distances.

\begin{figure}[b]
\begin{center}
\includegraphics[
width=3.5in
]
{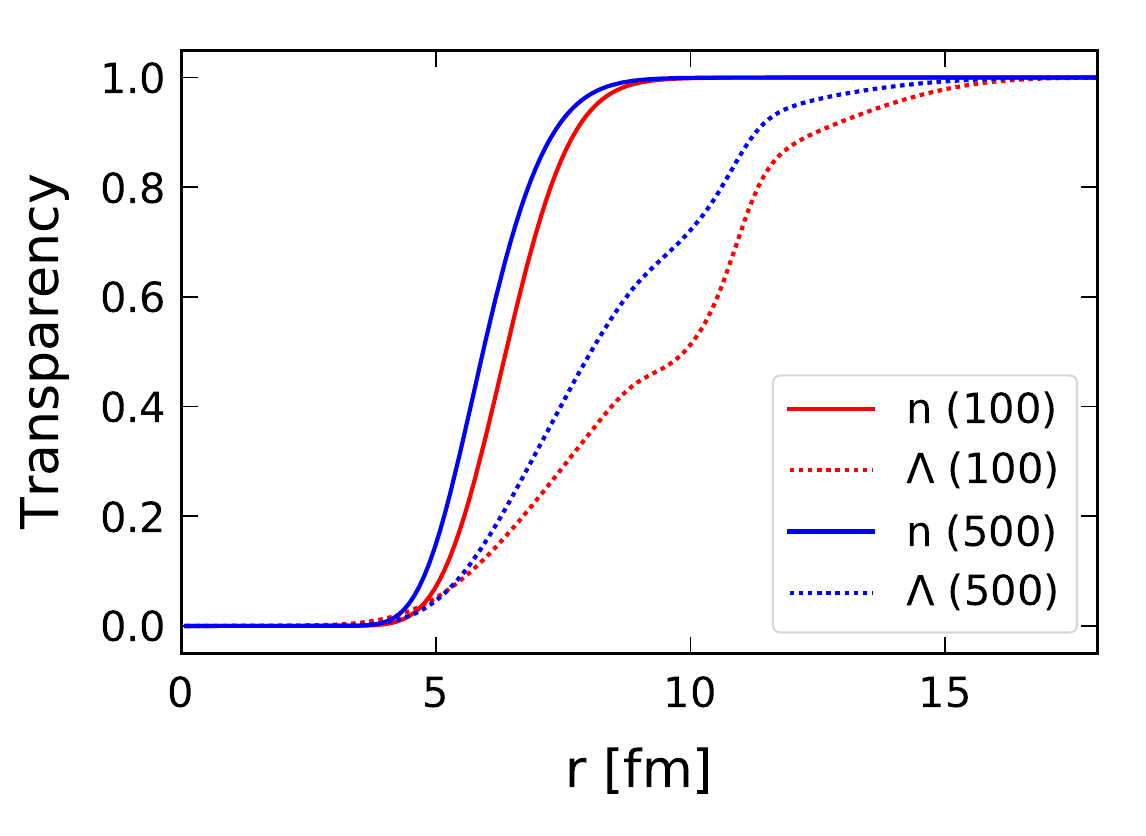}
\caption{Nucleon  and lambda transparencies in a collision of a hypertriton and a carbon target nucleus at 1500 MeV/nucleon. }
\label{transp}
\end{center}
\end{figure}

Once the particle densities are calculated, the transparency functions are obtained using Eq. \eqref{abrasionp2}. The numerical results are plotted in Figure \ref{transp} for the nucleon  and the lambda in a collision of a hypertriton and a carbon target at 1.5 GeV/nucleon. The transparency functions are straightforwardly related to the convolution of the distribution of each  incident particle with that of nucleons inside the carbon target. The nucleon density overlap of the hypertriton nucleons and those within the carbon target decay fast with increasing impact parameter. The corresponding transparency for nucleon removal is different than zero for impact parameters that are roughly the sum of the root mean square radii of their distributions inside the hypertriton and the carbon. The same behavior is observed for the knockout of the $\Lambda$ from  the hypertriton at small impact parameters. However, the transition from full opaqueness to full transparency is not so abrupt and displays changes in the slope. The slope changes arise when the $\Lambda$ density is sequentially probed from within the internal to the external part of the hypertriton where the $\Lambda$-deuteron wavefunction extends to large distances.


%

\end{document}